\documentclass[12pt]{article}
\usepackage{amssymb}
\usepackage{graphicx}
\usepackage{psfrag}
\usepackage{amsfonts}
\usepackage{epsfig}
\usepackage{rotate}
\usepackage{latexsym}

\newenvironment{appendletterA}
 {
  \typeout{ Starting Appendix \thesection }
  \setcounter{equation}{0}
  
 }{
  \typeout{Appendix done}
 }
\newenvironment{appendletterB}
 {
  \typeout{ Starting Appendix \thesection }
  \setcounter{equation}{0}
  
 }{
  \typeout{Appendix done}
 }

\newcommand{\sef}{\sin^2 \theta_{eff}^{lept}}
\newcommand{\ini}{\begin{equation}}
\newcommand{\fin}{\end{equation}}

\newcommand{\es}{s_{eff}^2}
\newcommand{\ses}{\sigma_{s_{eff,l}^2}}

\newcommand{\dr}{\Delta r}
\newcommand{\dre}{\Delta r_{eff}}

\newcommand{\dah}{\Delta \alpha_h^{(5)}}

\newcommand{\asmz}{\alpha_s \left( M_Z \right)}
\newcommand{\mhlb}{(M_H)_{L.B.}}
\newcommand{\bmath}{\begin{displaymath}}
\newcommand{\emath}{\end{displaymath}}
\newcommand{\bite}{\begin{itemize}}
\newcommand{\eite}{\end{itemize}}

\makeatletter
 \def\citenum#1{{\def\@cite##1##2{##1}\cite{#1}}}
\def\citea#1{\@cite{#1}{}}

\newcount\@tempcntc
\def\@citex[#1]#2{\if@filesw\immediate\write\@auxout{\string\citation{#2}}\fi
  \@tempcnta\z@\@tempcntb\m@ne\def\@citea{}\@cite{\@for\@citeb:=#2\do
    {\@ifundefined
       {b@\@citeb}{\@citeo\@tempcntb\m@ne\@citea\def\@citea{,}{\bf }\@warning
       {Citation `\@citeb' on page \thepage \space undefined}}%
    {\setbox\z@\hbox{\global\@tempcntc0\csname b@\@citeb\endcsname\relax}%
     \ifnum\@tempcntc=\z@ \@citeo\@tempcntb\m@ne
       \@citea\def\@citea{,}\hbox{\csname b@\@citeb\endcsname}%
     \else
      \advance\@tempcntb\@ne
      \ifnum\@tempcntb=\@tempcntc
      \else\advance\@tempcntb\m@ne\@citeo
      \@tempcnta\@tempcntc\@tempcntb\@tempcntc\fi\fi}}\@citeo}{#1}}
\def\@citeo{\ifnum\@tempcnta>\@tempcntb\else\@citea\def\@citea{,}%
  \ifnum\@tempcnta=\@tempcntb\the\@tempcnta\else
  {\advance\@tempcnta\@ne\ifnum\@tempcnta=\@tempcntb \else\def\@citea{--}\fi
    \advance\@tempcnta\m@ne\the\@tempcnta\@citea\the\@tempcntb}\fi\fi}
\makeatother

\begin{document}

\hyphenation{re-nor-ma-li-za-tion}

\begin{flushright}
NYU-TH/04-01-03\\
Freiburg-THEP 04/03\\
hep-ph/0401196
\end{flushright}

\vspace{0.5cm}
\begin{center}
\boldmath{\Large \bf Bounds on $M_W$, $M_t$, $\sef$}\unboldmath\\
\vspace{0.2cm}
\vspace{0.5cm}
{\large
A.~Ferroglia$^{a,}$\footnote{e-mail: andrea.ferroglia@physik.uni-freiburg.de},
G.~Ossola$^{b,}$\footnote{e-mail: giovanni.ossola@physics.nyu.edu},
and A.~Sirlin$^{b,}$\footnote{e-mail: alberto.sirlin@nyu.edu}}


\vspace{0.5cm}
$^a${\it Fakult\"at f\"ur Physik,  Universit\"at Freiburg,\\ D-79104 Freiburg, Germany\\ and\\
Institut f\"ur Theoretishe Theilchenphysik, Universit\"at
Karlsrhue,\\
D-76128 Karlsruhe, Germany} \\
\vspace{0.5cm}
$^b${\it Department of Physics, New York University,\\
4 Washington Place, New York, NY 10003, USA}
\end{center}
\bigskip
\bigskip
\bigskip

\begin{center}
\bf Abstract
\end{center}
{ \small Assuming that the Standard Model is correct and taking into account
the lower bound on $M_H$ from direct searches, we discuss
bounds on $M_W$, $M_{top}$, and $\sef$ at various confidence levels.
This permits to identify theoretically favored ranges for these important
parameters in the Standard Model framework, regardless of other observables.
As an illustration of possible future developments, a hypothetical benchmark
scenario, involving shifts $\lesssim 1 \sigma$ in the experimental 
central values, is discussed.
}

\newpage

\section{Introduction}

The general consensus at present is that the Standard Model (SM) gives
a very good description of a multitude of phenomena from atomic energies
up to the electroweak scale.
On the other hand, a fundamental pillar of the theory, the Higgs boson,
has not been found so far and some experimental observables put
sharp constraints on its mass.
This is particularly true of the $M_W$ measurement.
For instance, it was pointed out in Ref.~\cite{m1} that the 2002 average value
$(M_W)_{exp} = 80.451 \pm 0.033 \, \mbox{GeV}$, in conjunction with
$(M_t)_{exp} = 174.3 \pm 5.1 \, \mbox{GeV}$,  $\dah = 0.02761 \pm 0.00036$,
$\asmz = 0.118 \pm 0.002$, led to the prediction
$M_H = 23_{-23}^{+49}\, \mbox{GeV}$, and the 95\% C.L. upper bound
$M_H^{95} = 122 \, \mbox{GeV}$.
The first value is embarrassingly low relative to the 95\% C.L. lower bound
$\mhlb = 114.4\, \mbox{GeV}$ from direct searches \cite{m2}, while the second one is
only slightly larger. Since then the situation has changed significantly:
repeating this analysis with the new experimental value
$(M_W)_{exp} = 80.426 \pm 0.034 \, \mbox{GeV}$ \cite{m2}, we find, on the basis
of the simple formulae of Ref.~\cite{m1}, the predictions
$M_H = 45_{-36}^{+69}\, \mbox{GeV}$,  $M_H^{95} = 184 \, \mbox{GeV}$,
which are much less restrictive.
The predictions are further relaxed if one uses as inputs both
$(M_W)_{exp}$ and the current average value
$(\es)_{exp} = 0.23150 \pm 0.00016$ \cite{m2}, where $\es$ is an abbreviation
for $\sef$. This analysis leads to
\ini \label{e1}
M_H = 112_{-45}^{+69}\, \mbox{GeV};\quad M_H^{95} = 243 \, \mbox{GeV} ,
\fin
which are not far from the values currently derived from the global fit:
$M_H = 96_{-38}^{+60}\, \mbox{GeV}$,  $M_H^{95} = 219 \, \mbox{GeV}$ \cite{m2}.

There are three factors that single out the $M_W$ determination as
particularly significant: i) as illustrated in the above remarks, it places
sharp restrictions on $M_H$; ii) The LEP2 and collider measurements of
$M_W$ are in excellent agreement with $\chi^2/\mbox{Dof} = 0.3/1$;
iii) the relevant electroweak correction $\dr$ \cite{m3} has been fully
evaluated at the two-loop level \cite{m4}, an important theoretical achievement.

The aim of this paper is to derive bounds on $M_W$, $M_{top}$, and $\sef$
in the SM framework by comparing the experimental measurements of these 
three basic parameters
at various confidence levels with the theoretical functions $M_W=M_W(M_H,M_t)$
and $\es=\es(M_H,M_t)$, for fixed values of $M_H$.
The lower bound $\mhlb$ restricts the available parameter
space and this leads to bounds on  $M_W$, $M_{top}$, and $\sef$
that are significantly sharper than those derived from the experimental
measurements. Of course, this approach assumes the validity of the SM
and makes use of the $\mhlb$.
Thus, the derived bounds may be regarded as theoretically favored domains
for these parameters in the SM framework, regardless of other observables.
As such, they may suggest plausible ranges of variability in future, more
precise experimental determinations.

In Section~2 we examine the bounds derived from the theoretical
functions  $M_W=M_W(M_H,M_t)$ using the simple formulae from Ref.~\cite{m1},
as well as the new theoretical expressions presented in Ref.~\cite{cz}.
In Section~3  we extend the analysis to the functions $\es=\es(M_H,M_t)$
using the results of  Ref.~\cite{m1}.
In Section~4 we present the conclusions and, as an illustration of possible
future developments, we discuss a hypothetical benchmark scenario involving
shifts of $\lesssim 1 \sigma$ in the experimental central values.  
Appendix~A discusses the effect on the analysis
of the bounds due to the estimated errors in the $\dah$ determinations and
Appendix~B extends the analysis of Section~4 to the case of a ``theory driven''
calculation of $\dah$.

\boldmath\section{$M_W$ and $M_{t}$}\unboldmath

In this Section we compare the experimental values of $M_W$ and $M_t$
with the theoretical SM curves $M_W=M_W(M_H,M_t)$ for fixed values of $M_H$,
taking into account the lower bound $\mhlb$ on $M_H$ from the direct searches.
To simplify the analysis we take the restriction $M_H \ge 114.4\, \mbox{GeV}$
to be a sharp cutoff rather than a $95\%$
C.L. bound. The theoretical curves
depend also on $\dah$, the contribution from the first five quark flavors to
the running of $\alpha$ at the $M_Z$ scale.
We use as inputs $\dah = 0.02761 \pm 0.00036$ \cite{m6} and the ``theory
driven'' calculation  $\dah = 0.02747 \pm 0.00012$ \cite{m7}.
We also employ as inputs $\alpha$, $G_F$, $M_Z$, and  
$\asmz = 0.118 \pm 0.002$ \cite{m1,m2}.

\begin{figure}[t]
\centering \psfrag{m}{ {\small $M_t\, \left[ \mbox{GeV} \right]$}}
\psfrag{s}{{\footnotesize $M_W\, [\mbox{GeV}]$ }}
\resizebox{12cm}{7cm}{\includegraphics{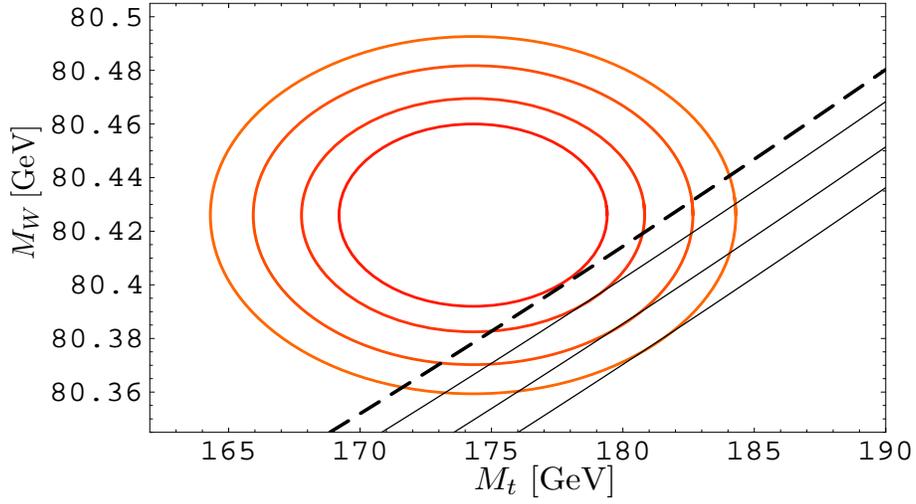}}
\caption{68\%, 80\%, 90\%, 95\% C.L. domains derived from
$(M_W)_{exp} = 80.426 \pm 0.034 \, \mbox{GeV}$ and $(M_t)_{exp} =
174.3 \pm 5.1 \, \mbox{GeV}$, together with the SM theoretical
curves $M_W(M_H,M_t)$ for $M_H = 114.4$ (dashed line), $139$,
$180$, $224\, \mbox{GeV}$ from Ref.~\cite{m1} with $\dah =
0.02761$. 
At a given C.L. the allowed region lies within the corresponding ellipse
and below the dashed boundary curve B.C..
 } \label{fig1}
\end{figure}

Fig.~\ref{fig1} shows the theoretical SM curves $M_W(M_H,M_t)$ for
$M_H = 114.4$ (dashed line), 139, 180, $224\, \mbox{GeV}$, $\dah =
0.02761$, and $\asmz = 0.118$, evaluated with the simple formulae
of Ref.~\cite{m1} in the effective scheme of renormalization
\cite{m1,m8}, as well as the 68\%, 80\%, 90\%, 95\% C.L. countours
derived from the current experimental values $(M_W)_{exp} = 80.426
\pm 0.034 \, \mbox{GeV}$, $(M_t)_{exp} = 174.3 \pm 5.1 \,
\mbox{GeV}$. An interesting feature is that the theoretical curves
are nearly linear over the range of $M_t$ values considered. At a
given C.L. the allowed region lies within the corresponding
ellipse and below the $M_H = 114.4\, \mbox{GeV}$ SM theoretical
curve (dashed line), which we call the boundary curve (B.C.). As
shown in Fig.~\ref{fig1}, the B.C. barely misses intersecting the
68\% C.L. ellipse, so that strictly speaking this region is not
allowed when the $\mhlb$ restriction is imposed. It turns out
that, to a good approximation, the maximum and minimum $M_W$ and
$M_t$ values in a given allowed region are determined by the
intersections of the B.C. with the associated ellipse. This
interesting feature can be understood by a glance at
Fig.~\ref{fig1}. The allowed $M_W$ and $M_t$ ranges determined by
such intersections are shown in Table~\ref{tab1} for the 80\%,
90\%, 95\% C.L. domains. As $M_H$ increases beyond $114.4\,
\mbox{GeV}$, the allowed ranges decrease in size. At a given C.L.
domain, the maximum allowed $M_H$ corresponds to the theoretical
curve  $M_W(M_H,M_t)$ that just touches the associated ellipse.
From Fig.~\ref{fig1} we can see that these values are $M_H \approx
139\, \mbox{GeV}$, $180\, \mbox{GeV}$, $224\, \mbox{GeV}$
corresponding to the 80\%, 90\%, 95\% C.L. domains.

In the above analysis $\dah$ has been kept fixed at the central value
 $\dah = 0.02761$. If it is allowed to vary according to
 $\dah = 0.02761 \pm 0.00036$, the analysis is somewhat more involved
(see Appendix~A). However, the conclusion is that the $M_W$, $M_t$
ranges reported in  Table~\ref{tab1} are at most affected by minor shifts.

Table~\ref{tab2} presents the  $M_W$, $M_t$
ranges evaluated with  $\dah = 0.02747$. In this case we see that there is
a very narrow window of compatibility with the 68\% C.L. domain.
Otherwise, the  $M_W$, $M_t$ ranges are very similar to those in Table~\ref{tab1}.
It is interesting to note that compatibility with the SM improves as
$\dah$ decreases.

Tables~\ref{tab3} and \ref{tab4} repeat the analysis of  Tables~\ref{tab1}
and~\ref{tab2} on the basis of the SM theoretical formulae presented
in Ref.\cite{cz}, which are based on a complete two-loop calculation
of $\dr$ \cite{m4} in the on-shell scheme of renormalization \cite{m3,m9}.
These Tables employ $\dah = 0.02761$ and  $\dah = 0.02747$, respectively.
Again  the 68\% C.L. domain is not compatible with the SM curves subject to
the $\mhlb$ restriction.
The allowed $M_W$, $M_t$ ranges in the  80\%, 90\%, 95\% C.L. domains
are similar but somewhat more restrictive than in  Tables~\ref{tab1}
and~\ref{tab2}. In particular, although the minimum $M_W$ values are
nearly the same, the maximum $M_W$ values are from 10 to 7 MeV smaller than in
Tables~\ref{tab1} and~\ref{tab2}.

In Table~\ref{tab5} we present the mid-points of the $M_W$ and
$M_t$ ranges in Table~\ref{tab2} with variations that cover the
full intervals. At a given C.L., these are compared with the
domains in $M_W$ and $M_t$ derived from $(M_W)_{exp} = 80.426 \pm
0.034 \, \mbox{GeV}$ and $(M_t)_{exp} = 174.3 \pm 5.1 \,
\mbox{GeV}$. As expected, the SM allowed $M_W$, $M_t$ ranges are
significantly restricted. To very good approximation, the
mid-points (80.402, 177.7) GeV are independent of the C.L. and are
shifted from the experimental central values $(M_W)_{exp}^c$ and
$(M_t)_{exp}^c$ by $\Delta M_W = -0.71 \sigma_{M_W}$ and $\Delta
M_t = +0.67 \sigma_{M_t}$.

In the case of Tables~\ref{tab1}, \ref{tab3}, and \ref{tab4}, to
very good approximation the mid-points are also independent of the
C.L. and are given by (80.401, 177.9) GeV, (80.397, 178.3) GeV,
and (80.398, 178.1) GeV, respectively. The largest shifts occur
for Table~\ref{tab3}, where the $M_W$ mid-point is $0.85
\sigma_{M_W}$ below $(M_W)_{exp}^c$ and the $M_t$ mid-point is
$0.78 \sigma_{M_t}$ above $(M_t)_{exp}^c$.

The theoretical formulae employed in this analysis are of course subject 
to errors associated with the truncation of the perturbative series and 
the QCD corrections. A relevant question is what is the effect of these
errors in the determination of the allowed $M_W$, $M_t$ ranges.
We give two specific examples concerning the 90\% C.L. domains in Tables~1
and~3. Using the estimated theoretical errors discussed in Ref.~\cite{m1},
we find that the 90\% C.L. ranges in Table~1 become $(60 \pm 5)$ MeV in 
$M_W$ and $({9.5}^{+0.8}_{-0.9})$ GeV in $M_t$. 
Using the estimated theoretical errors of Ref.~\cite{cz}, we  find that
the corresponding  90\% C.L. intervals in Table~3 become $({53}^{+3}_{-4})$ MeV in 
$M_W$ and $({8.8}^{+0.6}_{-0.8})$ GeV in $M_t$. 
Thus, the effect of the errors in the theoretical formulae is to change the size of
the allowed intervals by less then 10\%.
We also note that  these modifications are significantly smaller than the experimental
errors of $M_W$ and $M_t$ expected at TeV-LHC (cf.~Section~4).

\boldmath\section{$\sef$ and $M_{t}$}\unboldmath

On the experimental side, we consider two possibilities: the current world
average $(\sef)_{exp} = 0.23150 \pm 0.00016$ \cite{m2} and the average derived from the
leptonic observables  $(\sef)_{(l)} = 0.23113 \pm 0.00021$ \cite{m2}.
The difference between these values reflects the well-known dichotomy
between the leptonic and hadronic determinations,
which differ by $\approx 3 \sigma$.
On the theoretical side, the relevant electroweak correction is $\dre$ \cite{m8,m11,m12}.
Unlike $\dr$, it has not been fully evaluated at the two-loop level.
For this reason, we simply employ the formulae of Ref.\cite{m1} in the
effective scheme of renormalization. They contain two-loop electroweak effects
enhanced by powers $(M_t^2/M_W^2)^n$ ($n=1,2$), as well as QCD corrections.
For $M_t$, $\dah$, and $\asmz$ we employ the same inputs as in Section~2.

\begin{figure}[b]
\centering \psfrag{m}{ {\small $M_t\, \left[ \mbox{GeV} \right]$}}
\psfrag{s}{{\footnotesize $\es$ }}
\resizebox{12cm}{7cm}{\includegraphics{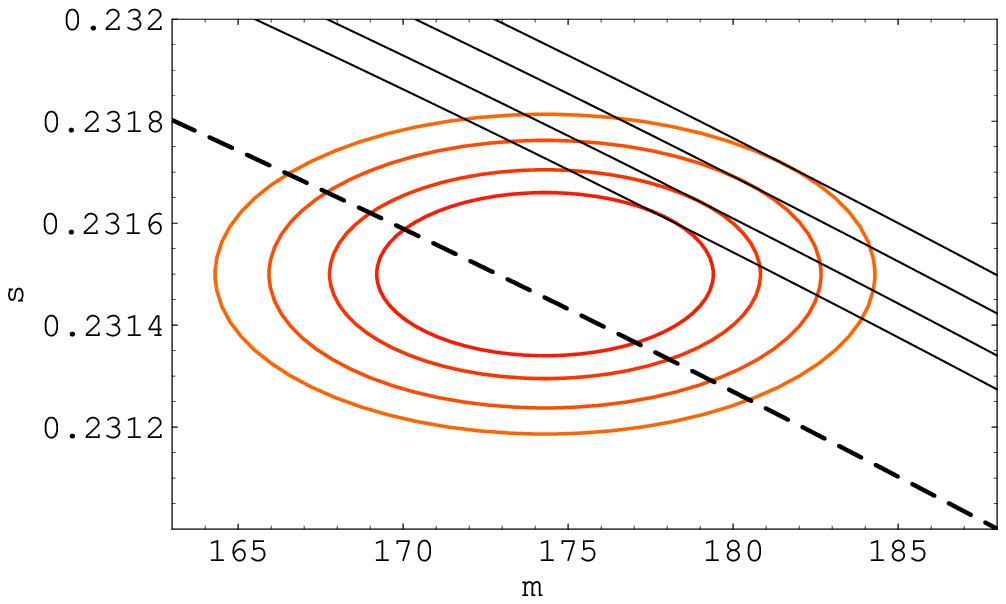}}
\caption{68\%, 80\%, 90\%, 95\% C.L. domains derived from $\sef =
0.23150 \pm 0.00016$ and $M_t = 174.3 \pm 5.1 \, \mbox{GeV}$,
together with the SM theoretical curves $\es(M_H,M_t)$ for $M_H =
114.4$ (dashed line), 193, 218, 253, $289\, \mbox{GeV}$ from
Ref.~\cite{m1} and $\dah = 0.02761$. The allowed regions lie above
the dashed B.C..} \label{fig2}
\end{figure}

Fig.~\ref{fig2} shows the 68\%, 80\%, 90\%, and 95\% C.L. domains
derived from the  world average $(\sef)_{exp} = 0.23150 \pm
0.00016$ and  $(M_t)_{exp} = 174.3 \pm 5.1 \, \mbox{GeV}$,  as
well as the SM theoretical curves $\es(M_H,M_t)$ for $M_H = 114.4$
(dashed line), $193$, $218$, $253$, $289\,\mbox{GeV}$,
 evaluated with
$\dah = 0.02761$ and $\asmz = 0.118$. At a given C.L. the allowed region lies within
the corresponding ellipse and above the B.C..
Since in this case the center of the ellipses lies in the allowed regions,
the situation is very different from that in Fig.~\ref{fig1}. In fact,
the reduction in parameter space is much less radical than in the ($M_W$,$M_t$)
analysis.
In particular, as shown in Fig.~\ref{fig2}, the maximum $\es$ and $M_t$ values are
not affected by the $(M_H)_{L.B.}$ restriction and the minimum values are increased
by relatively small amounts. 

In the case of the leptonic average  $(\sef)_{(l)} = 0.23113 \pm 0.00021$,
the situation is depicted in Fig.~\ref{fig3}. The allowed regions
lie again within the C.L. ellipses and above the B.C. (dashed line).
The 68\% C.L. domain is clearly forbidden. 

In the $(\sef)_{(l)}$, $M_t$ analysis, the effect of varying $\dah$
according to  $\dah = (\dah)^c \pm \sigma_{\Delta \alpha}$
is more pronounced than in the $M_W$,  $M_t$ case and,
accordingly, we have derived the allowed intervals using the $\chi^2$-analysis
discussed in Appendix A. They are shown in Tables~\ref{tab8}
and~\ref{tab9}, for $\dah = 0.02761 \pm 0.00036$ and
$\dah = 0.02747 \pm 0.00012$, respectively. To good approximation,
the mid-points are again independent of the C.L. and are given by 
(0.23129; 177.3 GeV) in Table~\ref{tab8}
and by (0.23129; 177.4 GeV) in Table~\ref{tab9}.
For a latter application in Section~4, we list also the mid-points of
the  $(\sef)_{(l)}$, $M_t$ ranges defined by the 
intersections of the $\dah = 0.02761$ and $\dah = 0.02747$ B.C.
with the C.L. ellipses. They are (0.23133; 178.1 GeV)  
and (0.23130; 177.5 GeV), respectively.
In the $\dah = 0.02761 \pm 0.00036$ case, they differ by relatively 
small amounts from the mid-points in Table~\ref{tab8},
mainly because the latter
take into account the effect of the  $\dah$ variation discussed before.
In the $\dah = 0.02747 \pm 0.00012$ case, because of the smallness of the error,
the effect is less significant and they almost coincide with the mid-points
in Table~\ref{tab9}.
The shifts of the B.C. mid-points 
(0.23133; 178.1 GeV) and (0.23130; 177.5 GeV)
from the experimental central values amount to 
($+0.95\, \sigma_{\es}$, $+0.75\, \sigma_{M_t}$) and
($+0.81\, \sigma_{\es}$, $+0.63\, \sigma_{M_t}$), respectively.

In analogy with the discussion at the end of Section~2, we now 
study the effect of the theoretical errors on the
determination of the $(\sef)_{(l)}$, $M_t$ ranges.
We consider, as an illustration, the 90\% C.L. intervals in Table~7. 
Taking into account the theoretical errors discussed in Ref~\cite{m1},
we find that these ranges become $0.00033 \pm 0.00005$ in $(\es)_l$ and
$({10.5}^{+1.0}_{-1.9})$ Gev in $M_t$. Thus, we see that the effect
of the theoretical errors is more pronounced than in the $M_W$, $M_t$ case,
changing the size of the allowed domains by less than 15\% in  $(\es)_l$
and 18\% in $M_t$.

\section{Discussion}

As is well known, global analyses of the electroweak observables in the
SM framework have frequently led to the derivation of preferred
domains in the $M_W$, $M_t$ plane at various C.L. \cite{m2,glob}. 
Although such studies are very valuable on general grounds,
the focus of the present paper is quite different. Specifically,
in Section~2 we have addressed the following question: given the
present experimental values of   $M_W$ and  $M_t$, and irrespective 
of other observables, what are the allowed ranges for these important
parameters in the SM framework when the $\mhlb$
is taken into account? In Section~3 the same question is addressed 
in the case of $\es$ and $M_t$. This approach conforms with the
idea that, aside from the global fits, it is also important to compare
the theory separately with the precise observables on which $M_H$ depends
most sensitively \cite{few}. In fact, it is in principle possible that
striking discrepancies of crucial observables and important
information may be blurred in the global analysis.
An obvious observation is that, if the SM is correct, the central values
of $M_W$, $M_t$, and $(\es)_l$ must approach the allowed regions
as the errors decrease, irrespective of what happens to other observables.
We also note that in cases such as $(M_W, M_t)$ and  $\left((\es)_l, M_t\right)$,
in which the central values lie well outside the allowed regions (cf. Fig.1 and Fig.3),
the bounds derived in this approach are significantly more restrictive than 
those obtained in the indirect global analysis (cf.~Fig.16.2 of Ref.~\cite{m2}).
Finally, it is worth pointing out that analyses of the kind carried out 
in this paper are particularly simple. For instance, the ellipses in Figs.1 and~3
are obtained from the experimental values by purely statistical means and the theory
essentially enters in the derivation of the theoretical curves.

In order to implement this approach, 
in Section 2 we have compared the experimental values for $M_W$ and $M_t$
at various C.L. with the SM theoretical curves  $M_W(M_H,M_t)$ for fixed $M_H$,
imposing the restriction $M_H \ge 114.4\, \mbox{GeV}$.
We have employed both the theoretical formulae of Ref.~\cite{m1} and Ref.~\cite{cz}
and considered two values of $\dah$.
As expected from the discussion in the Introduction, the
$M_W$ and $M_t$ ranges are significantly reduced in the SM theoretical framework
when the bound $M_H \ge 114.4\, \mbox{GeV}$ is taken into account.
Compatibility with the experimental 68\% C.L. region only
occurs in one of the alternatives we have considered and is at best marginal.
In the experimental 80\% C.L. domain, the current allowed $M_W$ and $M_t$ ranges
vary from $(M_W = 80.402 \pm 0.020\, ; \,  M_t = 177.7 \pm 3.1)\, \mbox{GeV}$ to
 $(M_W = 80.397  \pm 0.012\, ; \,  M_t = 178.3 \pm 2.0)\, \mbox{GeV}$,
depending on the value of $\dah$ and whether one employs the theoretical
expressions of Ref.~\cite{m1} or Ref.~\cite{cz}.

In order to belong to the allowed region, it is understood that
pairs of $M_W$ and $M_t$ values from these intervals should be
chosen so that they lie within the 80 \% C.L. domain.
At the 90\% and 95\% C.L., the allowed  $M_W$
and $M_t$ intervals are of course wider and can be read from
Tables~\ref{tab1}-\ref{tab4}; however, to a very good
approximation, in each Table the mid-points are independent of the
C.L..

The allowed $M_W$ and $M_t$ domains derived in this manner may be regarded as
theoretically favored in the SM framework when the $\mhlb$ is taken into account,
irrespective of other observables.
Qualitatively, they indicate that compatibility with the theory would improve if
 $(M_W)_{exp}^c$ would decrease and  $(M_t)_{exp}^c$ would increase.
Assuming the validity of the SM, the central values of $M_W$ and $M_t$
must approach  the allowed regions as the errors decrease.
Of course, the precise
end-point of this trajectory is not known, nor is it very clear what is the optimal C.L.
to select the allowed region. On the other hand,
the mid-points of the allowed regions provide natural representative examples.
The fact that to very good approximation they are independent of the C.L.
used in selecting the allowed regions (provided the C.L. are sufficiently large
that there are allowed regions), makes them particularly attractive benchmarks.
Therefore, we will consider, as an illustration of plausible future developments,
a hypothetical, but representative scenario in which the experimental central points
move to the mid-points of the current allowed intervals.

This would require a shift of $ -0.71$ to $-0.85 \sigma_{M_W}$ in  $(M_W)_{exp}^c$ and of
$+0.67$ to $0.78 \sigma_{M_t}$ in  $(M_t)_{exp}^c$.
It is interesting to note that a change in  $(M_W)_{exp}^c$ of the same direction and magnitude
occurred in the recent past: namely, the shift of  $(M_W)_{exp}^c = 80.451\, \mbox{GeV}$ to
 $(M_W)_{exp}^c = 80.426\, \mbox{GeV}$ represented a  $-0.76 \sigma_{M_W}$ effect.
Also, it is worthwhile to observe that the most precise $M_W$ measurement, the LEP2 determination
$(M_W)_{LEP2} = 80.412 \pm 0.042 \, \mbox{GeV}$ \cite{m2} has a central 
value that is significantly  closer
than  $(M_W)_{exp}^c$ to the mid-points mentioned above. Finally, 
there is a very recent preliminary value
$M_t = 180.1\, \pm 5.4\, \mbox{GeV}$ from the D0 collaboration 
that suggests that  $(M_t)_{exp}^c$ may
significantly increase in the near future \cite{d0}.

Assuming that the Higgs boson remains undiscovered, a natural
question is: what would be the $M_H$ prediction in this
hypothetical scenario? We use $\sigma_{M_W} = 15\, \mbox{MeV}$ and
$\sigma_{M_t} = 2\, \mbox{GeV}$, which are projected for Tev-LHC
\cite{hol},  $\dah = 0.02761 \pm 0.00036$, and $\asmz =0.118 \pm
0.002$. When the formulae of Ref.~\cite{m1} are employed, the
mid-points are (80.401, 177.9) GeV and we obtain the prediction
\ini \label{e2} M_H = 114_{-35}^{+46}\, \mbox{GeV};\quad M_H^{95}
= 195 \, \mbox{GeV} \, . \fin
Instead, using  Ref.~\cite{cz}, the mid-points are  (80.397,
178.3) GeV, and this leads to
\ini \label{e3} M_H = 114_{-37}^{+47}\, \mbox{GeV};\quad M_H^{95}
= 198 \, \mbox{GeV} \, . \fin
Eqs.~(\ref{e2},\ref{e3}) have been obtained without taking into account the
$\mhlb$. Including its effect we find that $M_H^{95}$ is shifted to 214
GeV in Eq.~(\ref{e2}) and 218 GeV in Eq.~(\ref{e3}).

In Section~3, we compared the experimental value for $\es$ and
$M_t$ at various C.L. with the SM theoretical curves
$\es=\es(M_H,M_t)$ for fixed $M_H$, taking into account the
$\mhlb$ restriction. Here we considered two alternatives: the
world average value for $\es$ and the average $(\es)_{(l)}$
derived from the leptonic observables. In the first case, there is
very good compatibility with $\mhlb$ and, in fact, the allowed
$\es$, $M_t$ intervals are only reduced by relatively small
amounts. In the second case, the 68\% C.L. is forbidden by the
$\mhlb$ and the allowed $(\es)_{(l)}$ and $M_t$ intervals are
significantly reduced. 
As in the case of the $M_W$, $M_t$
analysis, we may consider a hypothetical scenario in which the
experimental $(\es)_{(l)}$, $M_t$ central values move in the
future to representative points of the allowed region. 
In the $(\es)_{(l)}$, $M_t$ case, it is convenient to use as benchmarks
the mid-points of the ranges defined by the intersection of the B.C.
with the C.L. ellipses (cf. Section~3).
In order to
illustrate how this shift in the central values 
would affect the $M_H$ prediction, we assume
again $\sigma_{M_t} = 2\, \mbox{GeV}$, an error 
$\ses = 0.00001$ for $(\es)_{(l)}$, as projected for
$\es$ in the GigaZ application of the NLC,
and employ  $\dah = 0.02761 \pm 0.00036$, for
which the mid-points discussed in Section~3 are (0.23133; 178.1
GeV). These inputs lead to
\ini \label{e4} M_H = 115_{-29}^{+37}\, \mbox{GeV};\quad M_H^{95}
= 180 \, \mbox{GeV} \, . \fin
Including the effect of $\mhlb$, $M_H^{95}$ in Eq.~(\ref{e4}) 
is shifted to 196 GeV.

The central values in
Eqs.~(\ref{e2},\ref{e3},\ref{e4}) reflect the interesting feature 
that the mid-points are
close to the $M_H = 114.4\, \mbox{GeV}$ boundary curves.

\begin{figure}[t]
\centering \psfrag{m}{ {\small $M_t\, \left[ \mbox{GeV} \right]$}}
\psfrag{s}{{\footnotesize $\es$ }}
\resizebox{12cm}{7cm}{\includegraphics{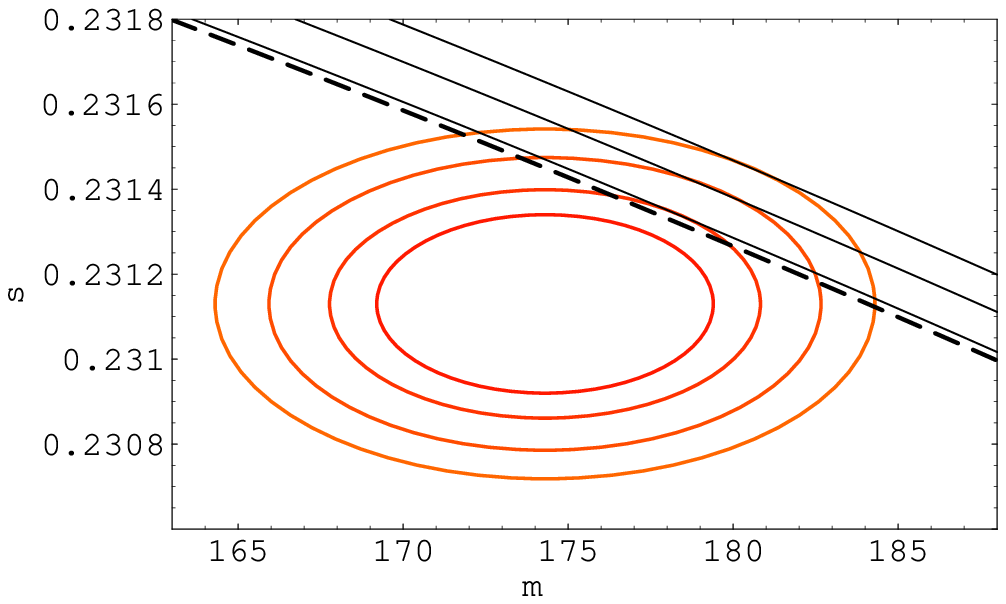}}
\caption{68\%, 80\%, 90\%, 95\% C.L. domains derived from
$(\sef)_{(l)} = 0.23113 \pm 0.00021$ and $M_t = 174.3 \pm 5.1 \,
\mbox{GeV}$, together with the SM theoretical curves
$\es(M_H,M_t)$ for $M_H = 114.4$ (dashed line), 119, 143, $169\,
\mbox{GeV}$ from Ref.~\cite{m1} and $\dah = 0.02761$. The allowed
regions lie above the dashed B.C..} \label{fig3}
\end{figure}

The fact that the benchmark scenario we have considered involves a
decrease in $(M_W)^c_{exp}$ and an increase in $(M_t)^c_{exp}$ and
$(\es)^c_{(l)}$ can be readily understood qualitatively by
observing the relative positions of the C.L. domains and the
allowed theoretical curves in Figs.~\ref{fig1} and \ref{fig3}.

If instead we assume that the current central values for $M_W$,
$M_t$, and  $(\es)_{(l)}$ remain unaltered while the errors
decrease to $\sigma_{M_W} = 15\, \mbox{MeV}$, $\sigma_{M_t} =
2\, \mbox{GeV}$, and $\ses = 0.00001$, 
the estimates of Eqs.~(\ref{e2},\ref{e3},\ref{e4}) are replaced by
\ini \label{e5} M_H = 45_{-18}^{+25}\, \mbox{GeV};\quad M_H^{95} =
90 \, \mbox{GeV} \, , \fin \ini \label{e6} M_H = 36_{-17}^{+23}\,
\mbox{GeV};\quad M_H^{95} = 79 \, \mbox{GeV} \, , \fin
\ini\label{e7} M_H = 59_{-16}^{+21}\, \mbox{GeV};\quad M_H^{95} =
96 \, \mbox{GeV} \, , \fin
respectively. Clearly, Eqs.~(\ref{e5},\ref{e6},\ref{e7}) would indicate a
sharp disagreement with the $\mhlb$!

An alternative possibility that would circumvent the
incompatibility of Eqs.~(\ref{e5},\ref{e6},\ref{e7}) would be an
increase of $(M_t)^c_{exp}$, with $(M_W)^c_{exp}$ and
$[(\es)_{(l)}]^c_{exp}$ kept fixed. As pointed out in
Ref.~\cite{d0}, such a shift would improve in general the
compatibility with the SM. This can be readily understood from
Figs.~\ref{fig1}-\ref{fig3}, since the C.L. ellipses would move
towards the allowed region.

If $(M_t)^c_{exp}$ increases by the current error $5.1 \,
\mbox{GeV}$, as discussed in Ref.~\cite{d0}, and we again employ
$\sigma_{M_W} = 15\, \mbox{MeV}$, $\sigma_{M_t} = 2 \,
\mbox{GeV}$, we would obtain from the $M_W$ input
\ini M_H = 87^{+38}_{-29} \,\mbox{GeV} \,;\qquad M_H^{95} = 155
\,\mbox{GeV} \, , \label{e8} \fin
using Ref.~\cite{m1}, and
\ini M_H = 74^{+36}_{-27} \, \mbox{GeV} \,;\qquad  M_H^{95} = 138
\,\mbox{GeV}\, , \label{e9} \fin
from Ref.~\cite{cz}. Employing the current value $(\es)_{(l)} = 0.23113$ and 
$\ses = 0.00001$, the result from the $(\es)_{(l)}$ input would be
\ini M_H = 83^{+29}_{-22} \,\mbox{GeV} \,; \qquad M_H^{95} = 134
\,\mbox{GeV} \, . \label{e10} \fin
Unlike Eq.~(\ref{e5},\ref{e6},\ref{e7}), Eq.~(\ref{e8},\ref{e9},\ref{e10}) are 
marginally compatible with the
$(M_H)_{L.B.}$. On the other hand,
Eqs.~(\ref{e8},\ref{e9},\ref{e10}), based on the $(M_t)^c_{exp} =
179.4 \, \mbox{GeV}$ assumption, are significantly more
restrictive than Eqs.~(\ref{e2},\ref{e3},\ref{e4}) corresponding
to the benchmark scenario discussed in this paper.

%
%
%
%

A qualitative difference between the two scenarios is that in
Eqs.~(\ref{e2},\ref{e3},\ref{e4}) the central experimental points reach the
allowed region, while this does not happen in
Eqs.~(\ref{e8},\ref{e9},\ref{e10}). In fact, in order to reach the allowed
region by varying $(M_t)^c_{exp}$ alone, one would need a shift of
$+7.5 \,\mbox{GeV}$ or $1.5$ times the current error if one
employs the $M_W$ input to predict $M_H$, and of $9.8
\,\mbox{GeV}$ or $1.9$ times the current error if one uses
$(s^2_{eff})_{(l)}$.

Throughout this Section we have employed  $\dah = 0.02761 \pm 0.00036$.
The corresponding $M_H$ estimates using the ``theory driven''
calculation $\dah$ $ = 0.02747 \pm 0.00012$ are presented in Appendix~B.


In summary, if the SM is correct, the experimental central values
should approach the allowed region as the errors shrink, 
regardless of other observables, and on
that basis the analysis of this paper suggests, as illustrated in
Figs.~\ref{fig1} and~\ref{fig3}, the possibility that
$(M_W)^c_{exp}$ will decrease, while $(M_t)^c_{exp}$ and
$(\es)^c_{(l)}$ will increase. In the hypothetical benchmark
illustration that we have described, all these changes are $< 1
\sigma$ in magnitude, so they are certainly not extreme. The fact
that shifts of this magnitude have recently occurred gives some
plausibility to this scenario. Thus, as the accuracy increases, it
will be very interesting to see whether the central values 
approach the allowed regions preferred by the SM, 
remain where they are, or move in the opposite
direction. In the last two cases a sharp disagreement with the SM
would emerge, thus providing strong evidence for new physics!


\section*{Acknowledgments}
The work of A.~S. was supported in part by NSF Grant PHY-0245068.
The work of A.~F. was supported by the DFG-Forschergruppe
\emph{``Quantenfeldtheorie, Computeralgebra und Monte-Carlo-Simulation''}.

\appendix
\section*{Appendix A}
\begin{appendletterA}

In the discussions of Section~2 and~3 we have kept $\dah$ fixed at
their central values. This has the advantage that the analysis is
particularly simple and can be readily illustrated in terms of
C.L. ellipses and the SM theoretical curves, as shown in the
figures. If $\dah$ is allowed to vary according to $\dah =
(\dah)^c \pm \sigma_{\Delta \alpha}$, the simplest procedure in
the $M_W$, $M_t$ case (Section~2) is to consider the $\chi^2$
function:
\bmath \chi^2 =
\frac{\left(M_W(M_H,M_t,\dah)-M_W^c\right)^2}{\sigma_{M_W}^2} +
\frac{\left(M_t - M_t^c\right)^2}{\sigma_{M_t}^2} \emath
\ini \label{a1} + \frac{\left(\dah -
(\dah)^c\right)^2}{\sigma_{\Delta \alpha}^2}\, , \fin
where $M_W^c$, $M_t^c$, and $(\dah)^c$ are the central values of
$(M_W)_{exp}$, $(M_t)_{exp}$ and $\dah$, respectively, and
$M_W(M_H,M_t,\dah)$ is the SM  theoretical curve that now depends
on $M_W$, $M_t$ and $\dah$. For fixed $\chi^2$ and $M_H$,
Eq.~(\ref{a1}) defines an implicit function $f(M_t,\dah) = \chi^2$
relating $M_t$ and $\dah$. Varying $\dah$ over an appropriate
finite interval, one finds numerically the range spanned by $M_t$.
Using  $M_W(M_H,M_t,\dah)$ one then obtains the domain of
variability of $M_W$. 
The resulting $M_W$, $M_t$ ranges agree, up to at most minor changes,  
with those presented in the Tables in Section~2.

In the $\es$, $M_t$ case, the effect of the $\dah =
(\dah)^c \pm \sigma_{\Delta \alpha}$ variation is more pronounced
and, accordingly, we have derived the allowed intervals in the 
Tables of Section~3 from a $\chi^2$-analysis analogous to that 
explained above.

\end{appendletterA}

\section*{Appendix B}
\begin{appendletterB}

In this Appendix we present the $M_H$ 
estimates discussed in Section~4 when one employs  the ``theory driven''
calculation $\dah = 0.02747 \pm 0.00012$ \cite{m7}, instead 
of $\dah = 0.02761 \pm 0.00036$ \cite{m6}.

We find that Eqs.~(\ref{e2},\ref{e3},\ref{e4}), corresponding
to our benchmark scenario with future projected errors, are
replaced by
\ini \label{b1} 
M_H = 115_{-34}^{+43}\, \mbox{GeV};\quad M_H^{95} = 191  \, \mbox{GeV} \, , 
\fin
\ini \label{b2} 
M_H = 114_{-35}^{+45}\, \mbox{GeV};\quad M_H^{95} = 194 \, \mbox{GeV} \, ,
\fin
\ini \label{b3} 
M_H = 115_{-17}^{+19}\, \mbox{GeV};\quad M_H^{95} = 148 \, \mbox{GeV} \, , 
\fin
respectively.

Eqs.~(\ref{e5},\ref{e6},\ref{e7}), corresponding to the current central values with 
future projected errors, are replaced by
\ini \label{b4} 
M_H = 48_{-18}^{+25}\, \mbox{GeV};\quad M_H^{95} = 92 \, \mbox{GeV} \, , 
\fin
\ini \label{b5} 
M_H = 38_{-17}^{+23}\, \mbox{GeV};\quad M_H^{95} = 80 \, \mbox{GeV} \, , 
\fin
\ini \label{b6} 
M_H = 65_{-10}^{+12}\, \mbox{GeV};\quad M_H^{95} = 86 \, \mbox{GeV} \, . 
\fin

Finally, instead of Eqs.~(\ref{e8},\ref{e9},\ref{e10}), corresponding to
the $(M_t)^c_{exp} = (174.3 + 5.1)\, \mbox{GeV}$ assumption, with 
current $(M_W)^c_{exp}$ and $[(\es)_{(l)}]^c_{exp}$ values, and 
future projected errors, we have
\ini \label{b7} 
M_H = 91_{-28}^{+38}\, \mbox{GeV};\quad M_H^{95} = 157 \, \mbox{GeV} \, , 
\fin
\ini \label{b8} 
M_H = 78_{-27}^{+35}\, \mbox{GeV};\quad M_H^{95} = 141 \, \mbox{GeV} \, ,
\fin
\ini \label{b9} 
M_H = 92_{-14}^{+16}\, \mbox{GeV};\quad M_H^{95} = 120  \, \mbox{GeV} \, . 
\fin

We see that the most significant change is between Eq.~(\ref{e4}) and 
Eq.~(\ref{b3}) derived from the $(\es)_{(l)}$ input in the benchmark scenario
discussed in this paper.

\end{appendletterB}

\newpage  

\begin{table}[!ht]
\begin{center}
\begin{tabular}{|c|c|c|}
\hline
&    &  \\
EFF scheme       &  range &  range \\
$\dah = 0.02761$ &  $M_W$ [GeV]    &  $M_t$ [GeV]    \\
&   &   \\
\hline
&  &  \\
80\% C.L. & 80.383 -- 80.419 & 175.0 -- 180.7  \\
&  &  \\
90\% C.L. & 80.371 -- 80.431 & 173.1 -- 182.6 \\
&  &  \\
95\% C.L. & 80.362 -- 80.441 & 171.6 -- 184.1  \\
&  &  \\
\hline
\end{tabular}
\caption{Comparison of the experimental values $(M_W)_{exp} =
80.426 \pm 0.034 \, \mbox{GeV}$ and $(M_t)_{exp} = 174.3 \pm 5.1
\, \mbox{GeV}$, at various C.L., with SM theoretical expressions
based on Ref.~\cite{m1} and $\dah = 0.02761$. The Table shows the
ranges for $M_W$ and $M_t$ that, according to the SM, are
compatible with the restriction $M_H \ge 114.4\, \mbox{GeV}$. In
order to belong to the allowed regions, pairs of  $M_W$ and $M_t$
values from these intervals should be chosen so that they lie
within the corresponding  C.L. domains.
%
Within each C.L. domain, the  $M_W$, $M_t$ ranges decrease as
$M_H$ increases from $114.4 \, \mbox{GeV}$. } \label{tab1}
\end{center}
\end{table}

\begin{table}[!ht]
\begin{center}
\begin{tabular}{|c|c|c|}
\hline
&    &  \\
EFF scheme       &  range &  range \\
$\dah = 0.02747$ &  $M_W$ [GeV]    &  $M_t$  [GeV]   \\
&   &   \\
\hline
&   &  \\
68\% C.L. & 80.396 -- 80.408 & 176.7 -- 178.7  \\
&  &  \\
80\% C.L. & 80.383 -- 80.422 & 174.6 -- 180.8  \\
&  &  \\
90\% C.L. & 80.371 -- 80.434 & 172.7 -- 182.6 \\
&  &  \\
95\% C.L. & 80.362 -- 80.443 & 171.3 -- 184.0  \\
&  &  \\
\hline
\end{tabular}
\caption{Same as in Table~\ref{tab1}, for $\dah = 0.02747$.}
\label{tab2}
\end{center}
\end{table}

\begin{table}[!ht]
\begin{center}
\begin{tabular}{|c|c|c|}
\hline
&    &  \\
Awramik et al. \cite{cz}  &  range &  range \\
$\dah = 0.02761$ &  $M_W$ [GeV]    &  $M_t$ [GeV]    \\
&   &   \\
\hline
&   &  \\
80\% C.L. & 80.385 -- 80.409 & 176.3 -- 180.3  \\
&  &  \\
90\% C.L. & 80.370 -- 80.423 & 173.9 -- 182.7 \\
&  &  \\
95\% C.L. & 80.361 -- 80.433 & 172.3 -- 184.2  \\
&  &  \\
\hline
\end{tabular}
\caption{Same as in Table~\ref{tab1}, with SM theoretical expressions from Ref.\cite{cz}.}
\label{tab3}
\end{center}
\end{table}

\begin{table}[!ht]
\begin{center}
\begin{tabular}{|c|c|c|}
\hline
&    &  \\
Awramik et al. \cite{cz}  &   range &   range \\
$\dah = 0.02747$ &  $M_W$ [GeV]     &  $M_t$ [GeV]    \\
&   &   \\
\hline
&   &   \\
80\% C.L. & 80.384 -- 80.413 & 175.7 -- 180.5  \\
&  &  \\
90\% C.L. & 80.371 -- 80.426 & 173.6 -- 182.7 \\
&  &  \\
95\% C.L. & 80.361 -- 80.435 & 172.0 -- 184.2  \\
&  &  \\
\hline
\end{tabular}
\caption{Same as in Table~\ref{tab3}, for $\dah = 0.02747$.}
\label{tab4}
\end{center}
\end{table}

\begin{table}[!ht]
\begin{center}
\begin{tabular}{|c||c|c||c|c|}
\hline
&   &    &   &  \\
 C.L.       &  $(M_W)_{exp}$   &  Allowed $M_W$   & Allowed $M_t$  & $(M_t)_{exp}$   \\
 &  [GeV]   &   [GeV]  &  [GeV]  & [GeV]\\
&   &    &   &\\
\hline
&   &    &  &  \\
68\% C.L. &  80.426 $\pm$ 0.034 &  80.402 $\pm$ 0.006 &    177.7 $\pm$ 1.0   &   174.3 $\pm$ 5.1    \\
&   &    &   &  \\
80\% C.L. &  80.426 $\pm$ 0.044 &  80.402 $\pm$ 0.020 &    177.7 $\pm$ 3.1   &   174.3 $\pm$ 6.5    \\
&   &    &  &  \\
90\% C.L. & 80.426 $\pm$ 0.056 &  80.402 $\pm$ 0.032  &    177.7 $\pm$ 5.0  &   174.3 $\pm$ 8.4    \\
&   &    &  &  \\
95\% C.L. &  80.426 $\pm$ 0.067 &  80.402 $\pm$ 0.041 &    177.7 $\pm$ 6.4   &   174.3 $\pm$ 10.0    \\
&   &    &  &  \\
\hline
\end{tabular}
\caption{Allowed $M_W$, $M_t$ ranges from Table~\ref{tab2}, expressed as mid-points and variations
covering the corresponding intervals. They are compared with the ranges extracted from
the experimental values.}
\label{tab5}
\end{center}
\end{table}



\begin{table}[!ht]
\begin{center}
\begin{tabular}{|c|c|c|}
\hline
&    &  \\
EFF scheme       &   range &   range \\
$\dah = 0.02761$ &  $\left(\sef \right)_l$  &  $M_t$  [GeV]   \\
&   &   \\
\hline
&   &  \\
80\% C.L. & 0.23119 -- 0.23139 & 174.7 -- 179.9   \\
&  &  \\
90\% C.L. & 0.23111 -- 0.23147 & 172.1 -- 182.6 \\
&  &  \\
95\% C.L. & 0.23105 -- 0.23153 & 170.4 -- 184.3  \\
&  &  \\
\hline
\end{tabular}
\caption{Comparison of the experimental values $(\es)_l = 0.23113 \pm
0.00021$ and $M_t = 174.3 \pm 5.1 \, \mbox{GeV}$ at various C.L.
with SM theoretical expressions from Ref.~\cite{m1} and $\dah =
0.02761 \pm 0.00036$. The Table shows the ranges for $\es$ and $M_t$ that,
according to the SM, are compatible with the restriction $M_H \ge
114.4\, \mbox{GeV}$. In order to belong to the allowed regions,
pairs of $\es$ and $M_t$ values from these intervals should be
chosen so that they lie within the corresponding  C.L. domains.
}
\label{tab8}
\end{center}
\end{table}

\begin{table}[!ht]
\begin{center}
\begin{tabular}{|c|c|c|}
\hline
&    &  \\
EFF scheme       &   range &   range \\
$\dah = 0.02747$ &  $\left(\sef \right)_l$    &  $M_t$  [GeV]   \\
&   & \\
\hline
&   &  \\
80\% C.L. & 0.23119 -- 0.23140 & 174.2 -- 180.6   \\
&  &  \\
90\% C.L. & 0.23113 -- 0.23146 & 172.2 -- 182.7 \\
&  &  \\
95\% C.L. & 0.23107 -- 0.23151 & 170.6 -- 184.2  \\
&  &  \\
\hline
\end{tabular}
\caption{Same as in Table~\ref{tab8}, for $\dah = 0.02747 \pm 0.00012$.
}
\label{tab9}
\end{center}
\end{table}


\begin{thebibliography}{99}

\bibitem{m1}
A.~Ferroglia, G.~Ossola, M.~Passera and A.~Sirlin,
Phys.\ Rev.\ D {\bf 65} (2002) 113002
[arXiv:hep-ph/0203224].

\bibitem{m2}
[LEP Collaboration],
arXiv:hep-ex/0312023.

\bibitem{m3}
A.~Sirlin, Phys. Rev. {\bf D22}  (1980) 971 and
Phys.\ Rev.\ {\bf D29} (1984) 89;\\
W.~J.~Marciano and A.~Sirlin, Phys.\ Rev.\ {\bf D22}  (1980) 2695.

\bibitem{m4}
M.~Awramik and M.~Czakon,
Phys.\ Rev.\ Lett.\  {\bf 89}, 241801 (2002)
[arXiv:hep-ph/0208113];
A.~Onishchenko and O.~Veretin,
Phys.\ Lett.\ B {\bf 551}, 111 (2003)
[arXiv:hep-ph/0209010];
M.~Awramik, M.~Czakon, A.~Onishchenko and O.~Veretin,
Phys.\ Rev.\ D {\bf 68}, 053004 (2003)
[arXiv:hep-ph/0209084];
M.~Awramik and M.~Czakon,
Phys.\ Lett.\ B {\bf 568}, 48 (2003)
[arXiv:hep-ph/0305248].




\bibitem{cz}
M.~Awramik, M.~Czakon, A.~Freitas and G.~Weiglein,
arXiv:hep-ph/0311148.

\bibitem{m6}
H.~Burkhardt and B.~Pietrzyk, LAPP-EXP-2001-03.

\bibitem{m7}
J.~F.~de Troconiz and F.~J.~Yndurain,
Phys.\ Rev.\ D {\bf 65}, 093002 (2002)
[arXiv:hep-ph/0107318].


\bibitem{m8}
A.~Ferroglia, G.~Ossola and A.~Sirlin,
Phys.\ Lett.\ B {\bf 507} (2001) 147
[arXiv:hep-ph/0103001], and
arXiv:hep-ph/0106094;
A.~Sirlin,
Nucl.\ Phys.\ Proc.\ Suppl.\  {\bf 116} (2003) 53
[arXiv:hep-ph/0210361].

\bibitem{m9}
K.~Aoki et al., Prog.\ Theor.\ Phys.\ Suppl.\ {\bf 73} (1982) 1;
M.~B\"ohm, H.~Spiesberger, and W.~Hollik, Fortsh.\ Phys.\ {\bf 34}
(1986) 687;
W.~Hollik, \emph{ibid. {\bf 38}} (1990) 165;
E.~Kraus, Annals Phys.\ {\bf 262} (1998) 155
[arXiv:hep-th/9709154];
P.~A.~Grassi, Nucl.\ Phys.\ {\bf B560} (1999) 499
[arXiv:hep-th/9908188].


\bibitem{m11}
B.~A.~Kniehl and A.~Sirlin,
Eur.\ Phys.\ J.\ C {\bf 16}, 635 (2000)
[arXiv:hep-ph/9907293].

\bibitem{m12}
A.~Sirlin,
in {\it Proc. of the 19th Intl. Symp. on Photon and Lepton Interactions at High Energy LP99 } ed. J.A. Jaros and M.E. Peskin,
Int.\ J.\ Mod.\ Phys.\ {\bf A15S1} (2000) 398
[eConf {\bf C990809} (2000) 398]
[arXiv:hep-ph/9912227];
A.~Sirlin,
J.\ Phys.\ G {\bf 29} (2003) 213
[arXiv:hep-ph/0209079].


\bibitem{glob}
J.~Erler,
AIP Conf.\ Proc.\  {\bf 670}, 227 (2003)
[arXiv:hep-ph/0212272].


\bibitem{few}
G.~Degrassi, P.~Gambino, M.~Passera and A.~Sirlin,
Phys.\ Lett.\ B {\bf 418}, 209 (1998)
[arXiv:hep-ph/9708311];
M.~S.~Chanowitz,
Phys.\ Rev.\ Lett.\  {\bf 80}, 2521 (1998)
[arXiv:hep-ph/9710308];
W.~J.~Marciano,
J.\ Phys.\ G {\bf 29}, 225 (2003).

\bibitem{d0}
P.~Gambino,
arXiv:hep-ph/0311257.

\bibitem{hol}
W.~Hollik,
J.\ Phys.\ G {\bf 29}, 131 (2003).



\end{thebibliography}
\end{document}